\documentclass[useAMS,usenatbib,usedcolumn]{mn2e}
\usepackage[figuresright]{rotating}
\usepackage{lscape}
\usepackage{graphics}
\usepackage{epsfig}
\usepackage{multirow}
\usepackage{bigdelim}
\usepackage{bigstrut}
\usepackage{amsmath}
\usepackage{amssymb}
\usepackage{lscape}
\usepackage{supertabular}
\usepackage{times}
\usepackage[T1]{fontenc}
\usepackage{aecompl}

\title[RAyMOND]{RAyMOND: An N-body and hydrodynamics code for MOND}
\author[G.~N. Candlish, R. Smith, M. Fellhauer]{G.~N. Candlish$^{1}$\thanks{E-mail: gcandlish@astro-udec.cl}, R. Smith$^{1}$, M. Fellhauer$^{1}$\\
$^{1}$Departamento de Astronom\'ia, Universidad de Concepci\'on, Casila 160-C, Concepci\'on, Chile}
\begin{document}

\maketitle

\begin{abstract}
The $\Lambda$CDM concordance cosmological model is supported by a wealth of observational evidence, particularly on large scales. At galactic scales, however, the model is poorly constrained and recent observations suggest a more complex behaviour in the dark sector than may be accommodated by a single cold dark matter component. Furthermore, a modification of the gravitational force in the very weak field regime may account for at least some of the phenomenology of dark matter. A well-known example of such an approach is MOdified Newtonian Dynamics (MOND). While this idea has proven remarkably successful in the context of stellar dynamics in individual galaxies, the effects of such a modification of gravity on galaxy interactions and environmental processes deserves further study. To explore this arena we modify the parallel adaptive mesh refinement code RAMSES to use two formulations of MOND. We implement both the fully non-linear aquadratic Lagrangian (AQUAL) formulation as well as the simpler quasi-linear formulation (QUMOND). The relevant modifications necessary for the Poisson solver in RAMSES are discussed in detail. Using idealised tests, in both serial and parallel runs, we demonstrate the effectiveness of the code.
\end{abstract}


\section{Introduction}
\label{intro}
For many years now the behaviour of dark matter has been intensively studied, both observationally and using simulations across a wide range of scales, leading to the development of the dominant $\Lambda$CDM concordance cosmological model. Multiple lines of observational evidence support the supposition that a cold dark matter component comprises around 27\% of the energy budget of the Universe. These include CMB observations \citep{Planck}, large-scale structure, gravitational lensing and cluster dynamics (see e.g. \cite{WeinbergCosmo} or \cite{PeeblesReview} for a more recent review and references). Nonetheless, due to the still mysterious nature of the dark matter itself, it is difficult to constrain the model sufficiently well to make concrete predictions at small scales. Therefore there remains several unresolved problems, such as the missing satellites problem, the cusp/core issue and the predicted existence of satellite halos that are ``too big to fail'' (see e.g. \cite{WeinbergProbs} for a recent review of these issues). It remains to be seen if improved handling of baryonic physics in cosmological simulations can resolve any of these problems. In addition, recent observations \citep{vpos,Ibata} have pointed to the existence of large thin rotating disks of satellite galaxies around both the Milky Way and Andromeda, something which is difficult to incorporate into the heirarchical formation mechanism of $\Lambda$CDM.

The puzzles of the dark sector have provoked the consideration of modifying gravity in an attempt to reconcile observations with theory. Multiple examples exist in the literature (see the comprehensive review of \cite{Padilla}). The majority of these theories are addressed towards the mystery of dark energy, rather than attempting to replace dark matter. One well-known exploration of the latter possibility, however, is the MOdified Newtonian Dynamics (MOND) first proposed by \cite{milgromMondOriginal}, in which the standard Newtonian gravitational force is enhanced once the accelerations drop below an empirically determined value of $a_0 \sim 1.2 \times 10^{-10}$~m/s$^2$. Development of this idea to a full relativistic formulation, which is necessary for the construction of a cosmological model, has been considered by \cite{MONDrelativistic1}, \cite{MONDrelativistic2} and \cite{MONDrelativistic3} among many others (see \cite{famaeymcgaugh} and references therein). At this stage, however, there is still no definitive relativistic theory that reproduces the MOND phenomenology and does not exhibit any pathological or unwelcome behaviour (see \cite{bruneton} for a discussion of such problems). Preliminary numerical investigations of cosmology in MOND \citep{knebe, llinares, angusCosmo} suggest that structure formation begins at an earlier epoch than in $\Lambda$CDM and proceeds more rapidly, possibly producing too much structure at late times. Completely consistent numerical simulations of cosmology in MOND (i.e. utilising a full relativistic formulation) are currently lacking, however, so no definitive statements can yet be made.

Given the unresolved problems of $\Lambda$CDM at small scales, the MOND paradigm remains an intriguing possibility for galaxy dynamics, at the very least as a phenomenological model of the behaviour of dark matter in galaxies. The success of the simple MOND scaling relation in modelling galaxy rotation curves, across a wide range of mass scales, is certainly suggestive. This is perhaps exemplified most clearly in the case of the baryonic Tully-Fisher relation \citep{mcgaughBTF}, which seems to imply that galaxy rotation curves are remarkably insensitive to the details of the dark matter structure in their halos. Indeed, this relation implies that there is a significant ``missing baryon'' problem in galaxy mass dark matter halos, while there is a similar such problem in galaxy clusters in MOND. Galaxy clusters may be considered as the environment in which MOND begins to show significant weaknesses as compared to $\Lambda$CDM, as even with the modification of gravity there is still a requirement for some kind of additional unseen matter in clusters. For an extensive review of the successes and failures of MOND, and possible future directions, see \cite{famaeymcgaugh}.

As MOND does not obviate the need for an additional unseen matter component at cluster scales and beyond, it is worthwhile asking the question of exactly how well MOND performs at galactic scales. It is not clear how the broader picture of galaxy dynamics behaves in a MOND universe, and whether it is consistent with observations. To investigate such questions we require numerical simulations using a MOND gravitational solver. Previous work has laid the groundwork for the development of such codes, in particular the pioneering work of \cite{brada}. In \cite{coombesTiret} a MOND N-body code was used to investigate the stability of MOND disks, as well as the behaviour of galaxy mergers. \cite{llinares} modified the cosmological N-body code AMIGA along similar lines. A publically available code known as NMODY (\citealp{nmody}) has been available for some time, and has been used to investigate various aspects of stellar dynamics in MOND, such as galaxy mergers and dynamical friction (\citealp{nipoti2007,nipoti2008}). This code works on a fixed spherical grid, however, making it cumbersome to perform simulations of systems that exhibit little symmetry.

To allow us to run simulations that cover wide ranges of length scales (as necessary to consider satellite galaxies orbiting around hosts, or galaxies falling into clusters, for example) and to model gas physics, we have chosen to modify the powerful N-body/hydrodynamics adaptive mesh refinement code RAMSES \citep{ramsespaper}. Our aim is to explore MONDian galactic dynamics in more physically realistic settings, involving physical processes such as galaxy mergers, ram-pressure stripping and tidal stripping. In this paper we will describe the code itself, and demonstrate the performance of the code using simple idealised tests. In future publications we will use the code to examine MONDian galaxy dynamics.

There are various possibilities for constructing a theory that exhibits MOND phenomenology in weak accelerations, with two broad classes that may be referred to as modified inertia or modified gravity (see \cite{famaeymcgaugh}). In this work we will consider the modified gravity theories, which involves a modification of the standard Poisson equation for the gravitational potential. The ``traditional'' formulation of MOND in this vein is that proposed in \cite{aqualmond}, referred to as the AQUAL formulation (due to the equation being derived from an aquadratic Lagrangian). The non-linearity of this equation limits the numerical techniques that may be used in a computational solver. The numerical scheme in RAMSES, however, is readily adaptable to accommodate such a non-linear equation.

A more recent formulation of MOND, often referred to as QUMOND, was presented in \cite{qumondpaper}. In this method, the non-linearity of the original MOND equation was reduced to a quasi-linearity, by maintaining the standard Poisson equation, and invoking the presence of an auxiliary acceleration field or, equivalently, an additional density component (often referred to as a ``phantom dark matter'' distribution). For computational purposes, this formulation therefore amounts to solving the Poisson equation twice, using the modified density distribution in the second step. As this uses the usual linear Poisson equation, standard numerical techniques may be applied for this version of MOND, albeit with an additional step of calculating the  density distribution. In the course of developing our code, \cite{PORpaper} published a description of their own modification of RAMSES to include a QUMOND solver. To facilitate a comparison with the AQUAL formulation we have also incorporated a QUMOND solver into our code.

One interesting consequence of introducing an acceleration scale in the modification of gravity is that the strong equivalence principle is no longer satisfied. This means that the centre-of-mass motion of a system has a direct effect on the internal accelerations of that system. This is known as the external field effect (EFE), and has been the subject of several investigations \citep{EFEpaper1, EFEpaper2, EFEpaper3, EFEpaper4}. This behaviour manifests itself in different ways in the QUMOND and AQUAL formulations: in the former the effect acts on the internal accelerations in a direction parallel to the Newtonian gravitational acceleration, while in the latter it is parallel to the MOND gravitational acceleration. In general, these two vectors may not be aligned, giving rise to differently oriented torques acting on the stellar system. Our code facilitates a comparison of the EFE in these two formulations.

The paper is organised as follows: in Section \ref{modram} we describe the modifications made to the RAMSES code to incorporate the two formulations of MOND; in Section \ref{codetests} we show the results of N-body tests used to verify that the code is capturing the MOND dynamics correctly; and finally we summarise and conclude in Section \ref{concl}.

\section{Modifications to RAMSES for the MOND calculation}
\label{modram}
The technical details of the RAMSES code may be found in \cite{ramsespaper}. In this section we will summarise the changes necessary to implement both the ``classical'' Bekenstein-Milgrom MOND theory given by \cite{aqualmond}, which we will refer to as AQUAL throughout this paper, and the quasi-linear formulation of \cite{qumondpaper}, which we will refer to as QUMOND.

RAMSES uses an iterative Gauss-Siedel solver accelerated with a multigrid scheme to solve the Newtonian Poisson equation. In the AQUAL formulation, we must solve a non-linear Poisson equation, given by
\begin{equation}
\label{aqualeq}
\nabla \cdot \left( \mu \left( \frac{| \nabla \phi |}{a_0} \right) \nabla \phi \right) = 4\pi \rho
\end{equation}
where $\mu(x)$ is the interpolation function which allows the transition between Newtonian and MONDian dynamics, and $a_0$ is the MOND acceleration parameter. For systems whose accelerations are below $a_0$, the gravitational force will be enhanced beyond the expected Newtonian force. The only constraints on $\mu$ are that it must satisfy the following limits in order to produce MONDian enhancement of the accelerations in the weak gravity regime, and recover Newtonian gravity in the strong gravity regime: $\mu(x) \to x$ for $x \ll 1$, and $\mu(x) \to 1$ for $x \gg 1$. Several interpolation functions have been discussed in the literature, and it is a trivial matter to implement any of them in our code. For the moment we choose the simplest function:
\begin{equation}
\mu(x) = \frac{x}{1+x}.
\end{equation}
With this choice of interpolation function it is known that MOND contradicts Solar System observations (\citealp{solarsystem}), requiring the use of a function which makes a more rapid transition from the MONDian to the Newtonian regime. We will investigate the effects of different choices of interpolation function in future work. We set $a_0 = 1.2 \times 10^{-10}$~m/s$^2$ throughout this paper. Note that this value is user-defined within an input parameter file and so may be easily changed if required.

\subsection{The modified Poisson solver}

\subsubsection{AQUAL}
\label{subsubsec:aqual}
\begin{figure}
\centering
\includegraphics[width=8.0cm]{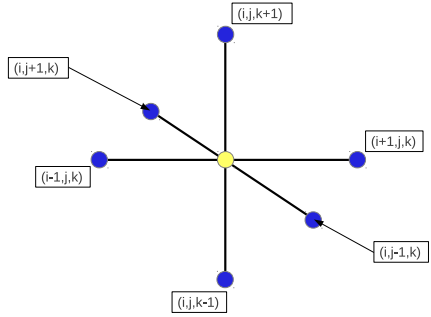}
\caption{6-point stencil used in the Gauss-Siedel iterative solver of standard RAMSES.}
\label{std_stencil}
\end{figure}

The Gauss-Siedel solver in RAMSES uses a standard 7-point stencil to update the value of $\phi$ at the central grid point by the average of $\phi$ on the six neighbouring grid points. This stencil is shown for comparison in Fig.~\ref{std_stencil}.

\begin{figure}
\centering
\includegraphics[width=8.0cm]{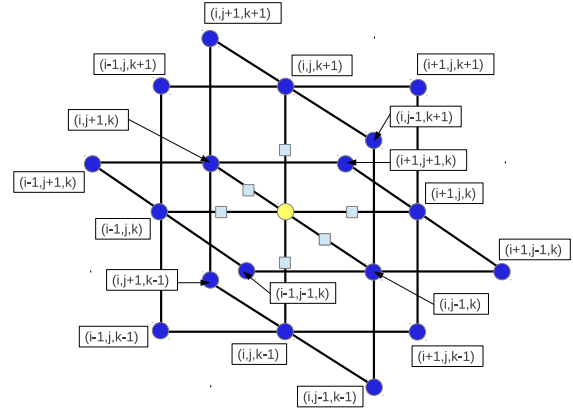}
\caption{18-point stencil used for the MOND solver. The squares indicate the locations where the $\mu$ function is evaluated.}
\label{mond_stencil}
\end{figure}

We will follow the procedure of \cite{brada}, \cite{coombesTiret} and \cite{llinares} by modifying the Gauss-Siedel solver used in RAMSES to use an extended stencil of points around each grid point in the calculation of $\phi$, as shown in Fig~\ref{mond_stencil}.
The updated value of $\phi$ at each grid point is now given by
\begin{equation}
\label{gsMOND}
\begin{split}
\phi_{i,j,k} &= \left( \mu_{i+1/2,j,k} \phi_{i+1,j,k} + \mu_{i-1/2,j,k} \phi_{i-1,j,k} \right. \\
            & \left. + \mu_{i,j-1/2,k} \phi_{i,j-1,k} + \mu_{i,j+1/2,k} \phi_{i,j+1,k} \right. \\
            & \left. + \mu_{i,j,k-1/2} \phi_{i,j,k-1} + \mu_{i,j,k+1/2} \phi_{i,j,k+1} \right. \\
            & \left. - 4\pi dx^2 \rho_{i,j,k} \right)/\sum_N \mu_N.
\end{split}
\end{equation}
The subscripts on the $\mu$ function denote that the function is to be evaluated midway between the grid points, as indicated by the squares in Fig.~\ref{mond_stencil}, and the denominator is the sum over $\mu$ at all 6 neighbouring midway points. The evaluation of $\mu$ requires a finite-difference calculation of the gradient of $\phi$ at the neighbouring grid points, and this requires the extended stencil shown in Fig.~\ref{mond_stencil}. As an example, the gradients of $\phi$ at the $i+1/2,j,k$ grid point (used in the evaluation of $\mu$ at that location) are calculated as follows:
\begin{equation}
\begin{split}
(\nabla \phi)^x_{i+1/2,j,k} &= \frac{\phi_{i+1,j,k} - \phi_{i,j,k}}{dx} \\
(\nabla \phi)^y_{i+1/2,j,k} &= \frac{\phi_{i,j+1,k} + \phi_{i+1,j+1,k} - \phi_{i,j-1,k} - \phi_{i-1,j-1,k}}{4dx} \\
(\nabla \phi)^z_{i+1/2,j,k} &= \frac{\phi_{i,j,k+1} + \phi_{i+1,j,k+1} - \phi_{i,j,k-1} - \phi_{i-1,j,k-1}}{4dx}.
\end{split}
\end{equation}
Similar expressions apply for the 5 other midway points at which $\mu$ is evaluated. Clearly the dependence of $\mu$ on $\phi$ leads to a $\phi$ dependence in the coefficients used in the Gauss-Siedel solver and thus a non-linear behaviour. We can easily see from Eq.~\ref{gsMOND} that we recover the Newtonian version of the Gauss-Siedel solver when $\mu = 1$, as is the case whenever all accelerations in the system are well above the MOND scale.

The extended stencil requires a modification of the RAMSES code as we now require $\phi$ values at \emph{diagonal} neighbours, such as $\phi_{i+1,j+1,k}$. Rather than substantially revise the data structure used in RAMSES to incorporate these new grid points as ``neighbours,'' we choose to use the existing linked list architecture to extract the points we require. Specifically, we find the neighbours of neighbours in order to reach the diagonal points.

At this point it is worth clarifying some terminology used in the RAMSES code. The computational mesh is structured in levels of refinement, the higher levels corresponding to more refinement (i.e. a higher resolution mesh). At each level, the mesh is organised into ``grids'' and ``cells.'' The cells are subdivisions of a grid, such that each grid contains 8 cells in three dimensions, 4 cells in two dimensions, and 2 cells in one dimension. A grid at one level corresponds to a cell at a lower level of refinement.

RAMSES uses a ``Fully Threaded Tree'' data structure \citep{Kolkov} in which the neighbouring \emph{grids} of a refined \emph{cell} are referenced with pointers for each cell. The neighbouring \emph{cell} is then reached from its associated parent grid. In order to reach the diagonal point, we ``leapfrog'' from the first neighbouring cell to the relevant neighbour of \emph{that} cell. In the event that the first neighbouring cell does not exist (i.e. the neighbouring \emph{grid} exists, as this is a cell at the next coarsest level, but that cell has not been refined further) then we move through \emph{two} levels of the mesh hierarchy (instead of just one) to find the diagonal point. In the event that the diagonal cell does not exist, we interpolate the $\phi$ value from the next coarsest level.

The multigrid algorithm utilised in RAMSES may be summarised in the following steps (more details may be found in \citealp{MGcitations}):
\begin{enumerate}
\item A simple initial guess is given for $\phi$.
\item The iterative solver updates $\phi$ on the fine mesh. This step is referred to as ``pre-smoothing'' in the context of the multigrid technique.
\item The residual (or error) is calculated on the fine mesh.
\item The values of the residual are transferred to the coarse mesh. This is known as the ``restriction'' step, and in RAMSES a simple average of the fine grid point values that surround a coarse grid point is assigned to the coarse grid point.
\item The \emph{correction} to the fine solution is calculated on the coarse mesh, using the coarse version of the iterative solver. Note that the correction is fixed to be zero in the boundary of the computational domain.
\item This coarse correction is then interpolated back to the fine mesh and added to the fine solution.
\item The iterative solver then updates the new corrected fine solution. This is known as ``post-smoothing.''
\end{enumerate}
This process may be represented symbolically as:
\begin{equation}
\begin{split}
& \phi_F^I \quad \text{smooth} \rightarrow \quad \overline{\phi}_F\\
& r_F = f_F - L_F \overline{\phi}_F \quad \text{Residual calculation}\\
& r_F \quad \text{restrict} \rightarrow \quad r_C \\
& L_C \epsilon_C = r_C \quad \text{Correction calculation}\\
& \epsilon_C \quad \text{interpolate} \rightarrow \quad \epsilon_F, \quad \overline{\phi}'_F = \overline{\phi}_F + \epsilon_F \\
& \overline{\phi}'_F \quad \text{smooth} \rightarrow \quad \overline{\overline{\phi}}_F
\end{split}
\end{equation}
where the $C$ and $F$ subscripts denote quantities on the fine and coarse meshes respectively. The linear matrix operator on the fine/coarse mesh $L_{F/C}$ is, of course, the finite-difference version of the differential operator $\nabla^2$ in the Poisson equation. The right-hand side of the Poisson equation is denoted by $f_F$, while the residual is $r_{F/C}$. It is important to note that it is the \emph{correction} to the fine solution that is solved for on the coarse mesh, rather than the coarse solution itself.

For the non-linear AQUAL formulation of MOND we must modify the above multigrid algorithm, in addition to modifying the Gauss-Siedel solver. Several approaches are possible: we choose the Full Approximation Storage scheme (for details see e.g. \cite{MGcitations}). In essence this requires that we calculate the full solution to Eq.~\ref{aqualeq} on all \emph{coarse} levels as well. This contrasts with the linear (Newtonian) case where only the correction is calculated on the coarse levels.

For the non-linear equation, the multigrid scheme now proceeds as follows:
\begin{equation}
\begin{split}
& \phi_F^I \quad \text{smooth} \rightarrow \quad \overline{\phi}_F, \quad \overline{\phi}_F \quad \text{restrict} \rightarrow \quad \overline{\phi}_C \\
& r_F = f_F - N_F(\overline{\phi}_F) \overline{\phi}_F \\
& r_F \quad \text{restrict} \rightarrow \quad r_C \\
& N_C(\psi_C) \psi_C = r_C + N_C(\overline{\phi}_C) \overline{\phi}_C \\
& \epsilon_C = \psi_C - \overline{\phi}_C \\
& \epsilon_C \quad \text{interpolate} \rightarrow \quad \epsilon_F, \quad \overline{\phi}'_F = \overline{\phi}_F + \epsilon_F \\
& \overline{\phi}'_F \quad \text{smooth} \rightarrow \quad \overline{\overline{\phi}}_F.
\end{split}
\end{equation}
The two main differences with the linear case are that we must restrict the initial fine solution after smoothing, and that we calculate the full coarse solution $\psi_C$, using the restricted fine solution, rather than just the correction. The restricted fine solution is then subtracted from the coarse solution to find the correction. As for the linear Newtonian case, we recursively apply this algorithm using the various levels of refinement.

\subsubsection{QUMOND}
\label{subsubsec:qumond}
The QUMOND version of the code does not require any modifications to the Gauss-Siedel solver, as this formulation involves solving the standard Poisson equation, albeit twice, and with a modified density field the second time. The calculation of the density field, however, makes use of the extended stencil. We must solve the following elliptic differential equation for the so-called ``phantom dark matter'' density of QUMOND:
\begin{equation}
\label{qumondeq}
\rho' = \frac{1}{4\pi G} \nabla \cdot \left( \tilde{\nu} \left( \frac{|\nabla \phi_N|}{a_0} \right) \nabla \phi_N \right)
\end{equation}
where the Newtonian potential $\phi_N$ is determined from a standard pass through the RAMSES multigrid algorithm. The MOND interpolation function used in this formulation is closely related to the inverse of that used in the AQUAL formulation, satisfying the limits $\tilde{\nu}(y) \to 0$ when $y \gg 1$ and $\tilde{\nu}(y) \to y^{-1/2}$ when $y \ll 1$, where in this case the argument is the ratio of the \emph{Newtonian} gravitational acceleration to the MOND scale. The $\tilde{\nu}$ function corresponding to our chosen $\mu$ function is
\begin{equation}
\frac{1}{2}\sqrt{1 + \frac{4}{y}} - \frac{1}{2}.
\end{equation}
This is related to $\nu$, the inverse of the $\mu$ function, by $\nu = \tilde{\nu} + 1$.

We can see that the functional form of Eq.~\ref{qumondeq} is very similar to Eq.~\ref{aqualeq}, and so we use the finite-difference stencil described for the AQUAL formulation to solve this equation. The difference is that we can determine $\rho'$ at each grid point using Eq.~\ref{qumondeq} immediately: we do not need to use an iterative scheme to converge to the result. After the additional density contribution has been calculated, we add this to the real density field of the system, and solve the standard Poisson equation a second time to determine the MOND potential.

This formulation does not require any modifications to the multigrid scheme. We must, however, make minor modifications to the sequence of execution through a time step used by RAMSES in order to solve the Poisson equation with the appropriate density field. In addition, we must introduce new arrays to store the MONDian potential and the ``phantom dark matter'' density, increasing the memory usage of the code.

\subsection{Dealing with the boundaries}
For points that lie outside the full computational domain, RAMSES uses a fixed analytic value of the gravitational potential to provide a boundary value. The default is to use $\phi = M/r$ (in three dimensions, with $G=1$): the potential sourced by a point mass at the origin, where the mass is equal to the total mass in the simulation. This may be easily modified to use other choices.

The situation is more complicated in the case of boundaries for an adaptive refined level, which typically is smaller than the full computational domain. In this case the potential required at the boundary is interpolated from the next lower level (i.e. the first ``coarse'' level before the current level). This interpolated value is then used to modify the right-hand side of the Poisson equation that is used in the solver. The precise location of the boundary at any particular level is determined using a masking procedure. Further details may be found in \cite{bndPaper}.

As for the Newtonian code, we must supply an analytic estimate of the potential at the domain boundary. This is done for both the AQUAL and QUMOND formulations using the asymptotic behaviour of the MOND potential sourced by an isolated system (\citealp{famaeymcgaugh}):
\begin{equation}
\phi(r) = \sqrt{GMa_0} \ln(r).
\end{equation}

The treatment of boundaries on adaptive mesh refinement (AMR) levels is unchanged in the QUMOND formulation. We must, however, modify this scheme for the AQUAL formulation, for the following reason. Firstly, recall that there are two mesh heirarchies in RAMSES: one is the set of levels used in the multigrid acceleration of the iterative solver, the other is the set of AMR levels used to obtain high resolution in regions of interest. The first set of levels is always a sub-set of the second set of levels, in the sense that the Poisson solver determines $\phi$ on each level, using all coarser levels for the multigrid acceleration. In the standard Newtonian case, the iterative solver determines the \emph{correction} on all the coarse levels in the multigrid scheme, not the full coarse solution. Therefore, the boundary condition on all the coarse levels is simply $\epsilon_C = 0$, using the notation of Section~\ref{subsubsec:aqual}. Thus the boundary treatment used in standard RAMSES is adapted to this scheme, using a masking procedure and a modification of the right-hand side of the Poisson equation to ensure that this boundary condition is applied.

In our non-linear solver, we cannot alter the right-hand side of Eq.~\ref{aqualeq} with $\phi$-dependent terms. Therefore, for each fine AMR level, we choose to find the boundary cells at the start of the multigrid algorithm, determine the required $\phi$ values in those cells by interpolation from the next coarsest level, and store these in an array for use in the Gauss-Siedel solver. This introduces some minor additional memory overhead to the code.

\subsection{Parallel computations}
The modifications made to the multigrid algorithm for the AQUAL formulation require some minor modifications to the parallelisation in RAMSES. The Full Approximation Storage scheme implies that we must communicate the fine resolution solution for $\phi$ to all the virtual boundaries of the CPUs. This is easily done using an additional call to the relevant communication subroutine already implemented in RAMSES. We have run tests using both the serial and parallel versions of the code, and the results agree exactly.

\section{Code tests}
\label{codetests}

\subsection{Analytic comparison}

\label{subsec:tests_analytic}

\begin{figure*}
\begin{tabular}{cc}
\includegraphics[width=7.5cm]{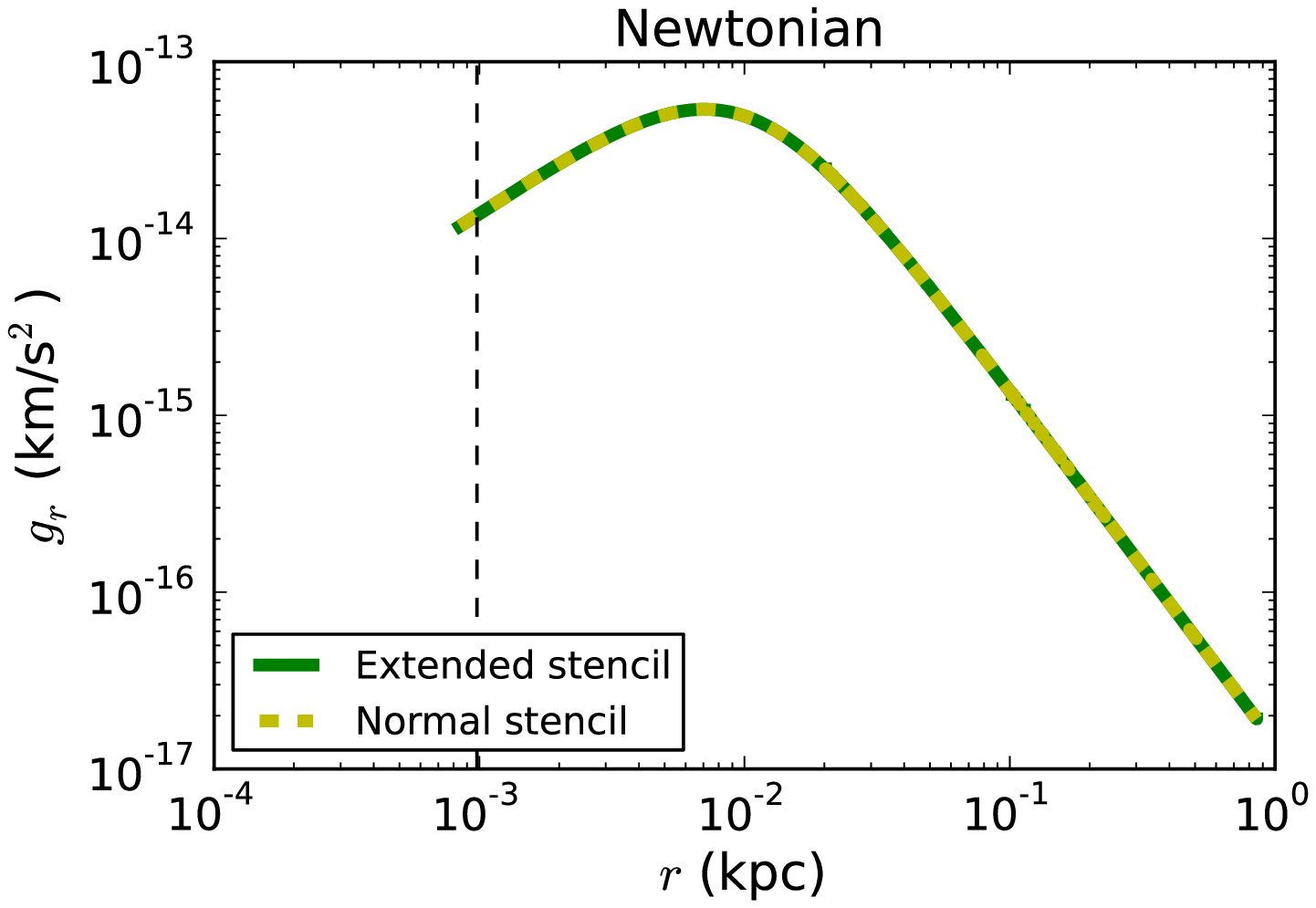} & \includegraphics[width=7.5cm]{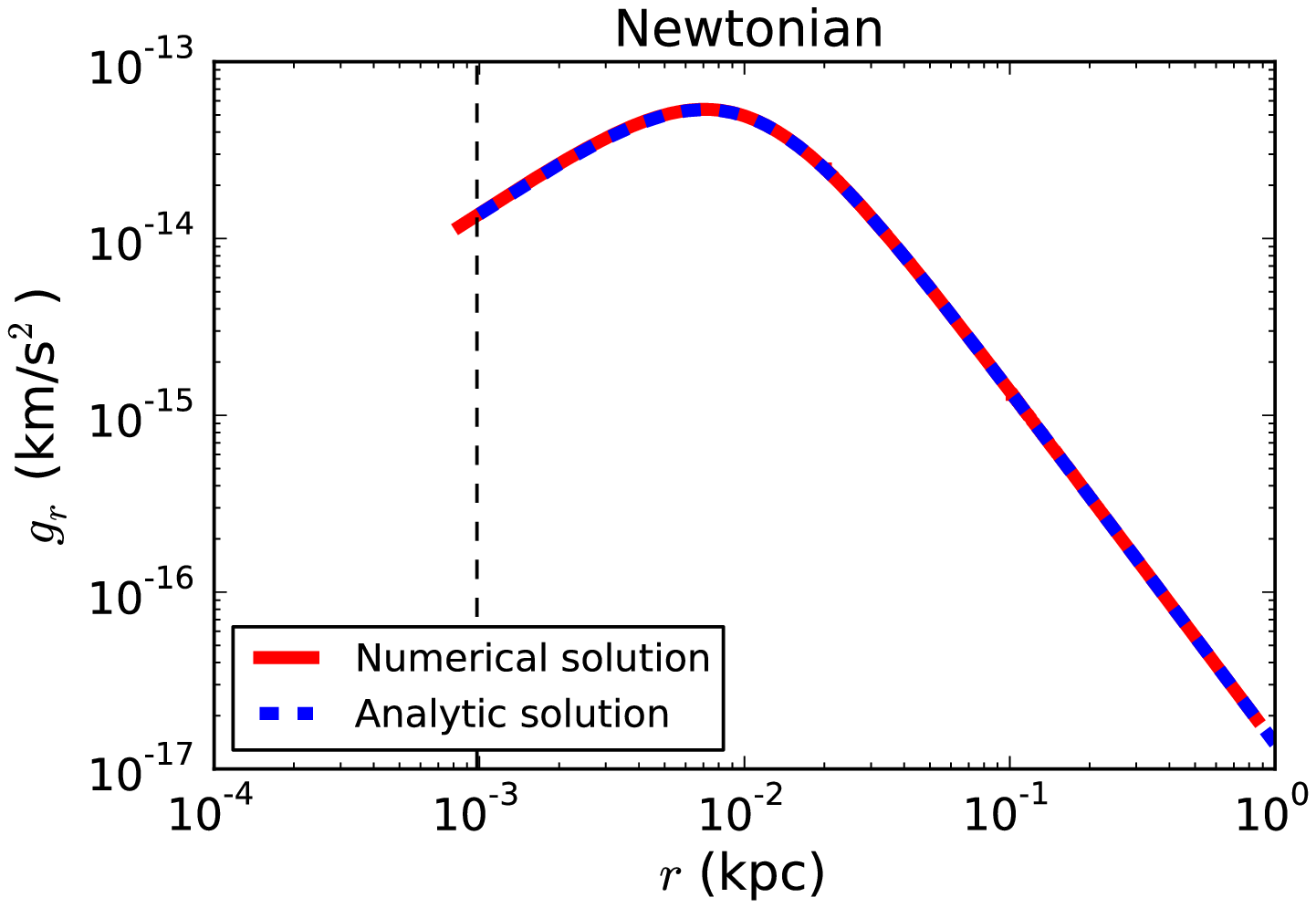}
\end{tabular}
\caption{The left-hand plot shows the gravitational acceleration calculated from a Newtonian analytic Plummer density using the standard 6-point stencil and the 18-point stencil of the modified solver. The right-hand plot shows the comparison of this numerically calculated gravitational acceleration with the analytic result. The vertical dashed line shows the resolution limit.}
\label{newtonised}
\end{figure*}

\begin{figure*}
\begin{tabular}{cc}
\includegraphics[width=7.5cm]{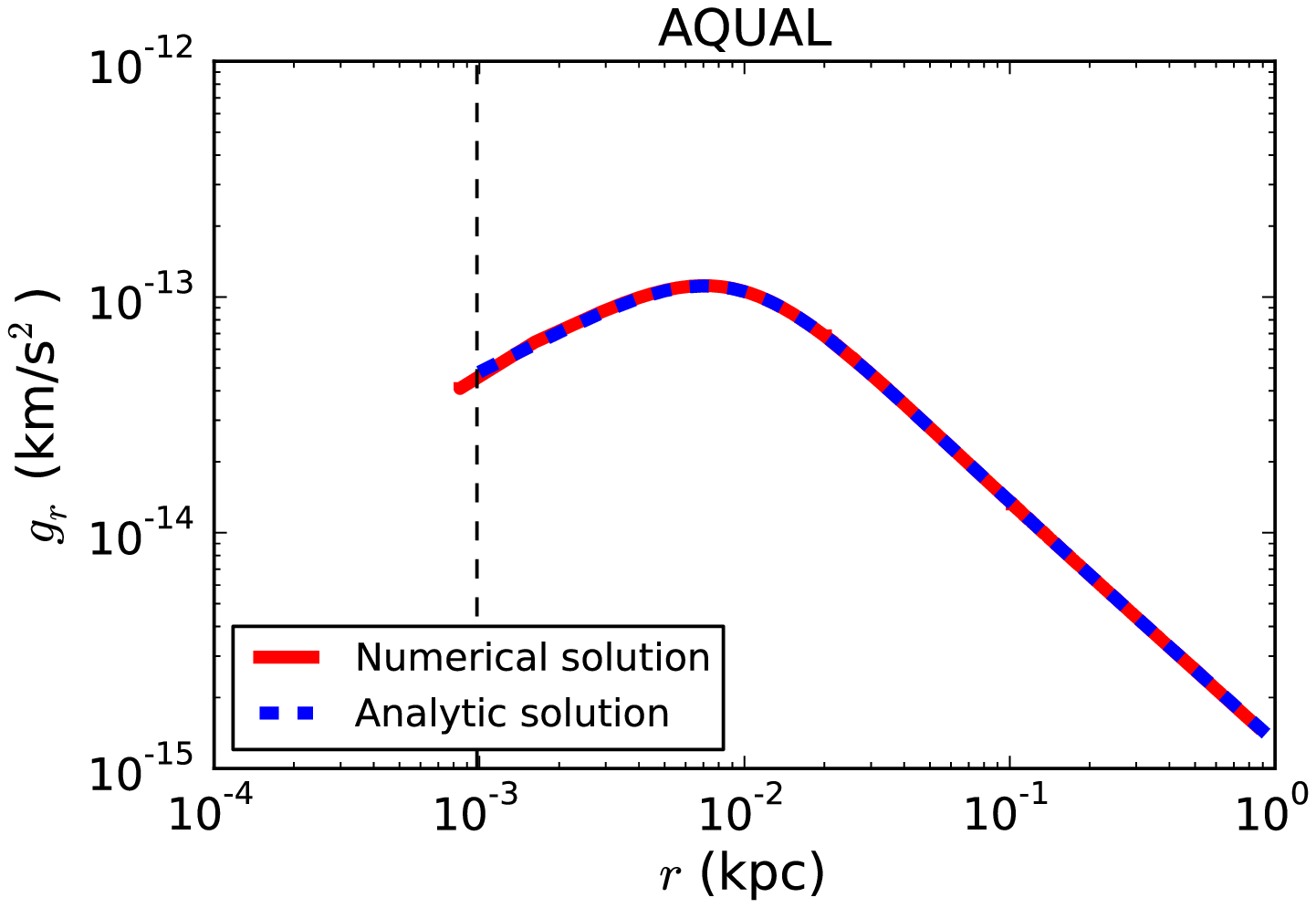} & \includegraphics[width=7.5cm]{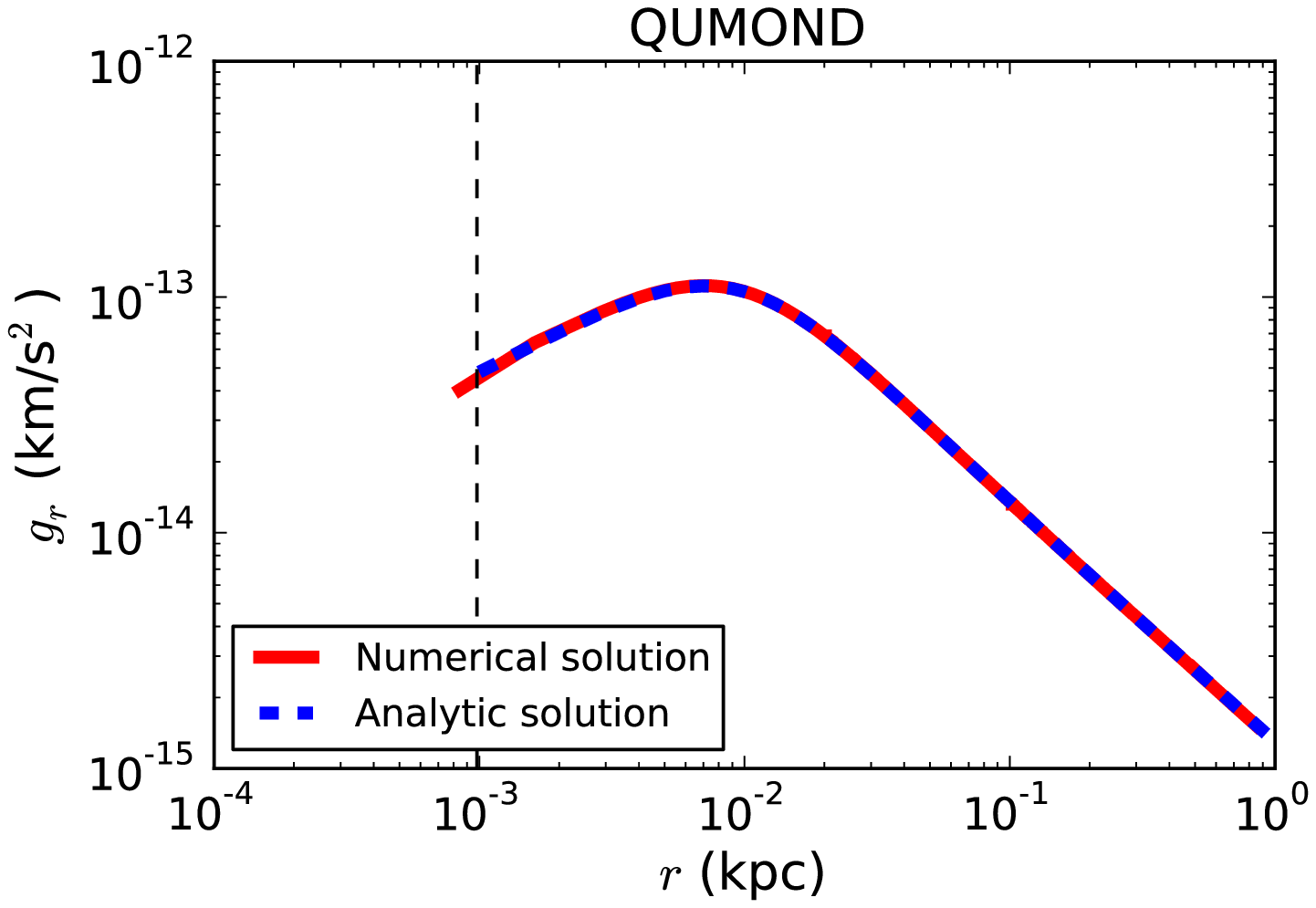}
\end{tabular}
\caption{Comparing the analytic radial gravitational acceleration in a Plummer sphere with the numerical solutions in the MOND cases. The vertical dashed line is the resolution limit.}
\label{analytictests}
\end{figure*}

While analytic MOND potentials are generally difficult to obtain, in the case of spherical symmetry the MOND gravitational acceleration $\vec{g}_M$ is straightforwardly related to the Newtonian acceleration $\vec{g}_N$ using the scaling relation
\begin{equation}
\label{eq:mondscaling}
\vec{g}_M = \nu(|\vec{g}_M|/a_0)\vec{g}_N,
\end{equation}
where we have used the inverse interpolating function $\nu$ discussed in Section~\ref{subsubsec:qumond} (not to be confused with $\tilde{\nu}$), as we can easily obtain the Newtonian acceleration analytically. Using this relationship we can test our code by including an analytic density distribution $\rho$ describing a spherically symmetric system, and running the solver for one time step. For our analytic density distribution we choose a Plummer profile:
\begin{equation}
\rho = \frac{3M_{pl}}{4\pi r_{pl}^3} \left( 1 + \frac{r^2}{r_{pl}^2} \right)^{-5/2}
\end{equation}
with the parameters $r_{pl} = 10$~pc and $M_{pl} = 10^5M_{\odot}$. The Newtonian radial gravitational acceleration for this profile is
\begin{equation}
g_r^N = \frac{GMr}{(r^2 + r_{pl}^2)^{3/2}},
\end{equation}
and the MONDian acceleration may be calculated from Eq.~\ref{eq:mondscaling}. The simulation box is $1$~kpc, and the computational domain is fully refined to level 6, giving a minimum resolution of $15.63$~pc, with AMR refinement to level 10, giving a maximum resolution of $0.98$~pc. The AMR refinement is specified using the geometric criteria of RAMSES, where we specify refinement within spherical regions centred at the origin, with diameters of $0.3, 0.15, 0.1, 0.05$~kpc.

The results are shown in Fig.~\ref{analytictests}. The vertical dashed line in the figure indicates the cell size of our highest resolution grid. We can see that the numerical solution agrees extremely well with the analytic solution. It is worth noting that the AQUAL and QUMOND formulations agree precisely in this spherically symmetric system.

Throughout the paper we use the standard RAMSES (i.e. the usual 6-point stencil) for Newtonian comparisons. As a test of our modifications for the extended stencil, we run an analytic test as above, but using the AQUAL code with $a_0 \sim 10^{-31}$~m/s$^2$. Setting the MOND scale to such a small value ensures that the system is entirely Newtonian, and we should therefore recover the result from standard RAMSES. The comparison using the analytic test is shown in Fig.~\ref{newtonised}, where we can see that the numerical solution from the AQUAL code for a fully Newtonian system is identical to that of the standard 6-point stencil RAMSES.

\subsection{N-body tests}

\subsubsection{Isolated Plummer spheres}
\label{subsec:tests_isoPlummer}
We now test the code using a live N-body system, consisting of a Plummer sphere with the same parameters as in Section~\ref{subsec:tests_analytic}, i.e. $r_{pl} = 10$~pc and $M_{pl} = 10^5M_{\odot}$, in a simulation box with length $1$~kpc. Again, the minimum level of the mesh is level 6, with refinement down to level 10, corresponding to minimum and maximum resolutions of $15.63$~pc and $0.98$~pc. Therefore, the Plummer radius corresponds to approximately 10 high resolution cell widths. The chosen mass and scale radius ensure this Plummer sphere is well inside the MONDian regime.

The initial conditions are generated using a setup code that performs a numerical integration of the one-dimensional Jeans equation to determine the radial velocity dispersion as a function of radius:
\begin{equation}
\label{1djeans}
\sigma_r^2(r) = \frac{1}{\rho(r)} \int_r^{\infty} dr' \rho(r') \frac{d\phi(r')}{dr'}.
\end{equation}
This velocity dispersion is then used to randomly assign isotropic velocities. Due to the spherical symmetry and the fact that this calculation only requires the radial gravitational acceleration, we can use Eq.~\ref{eq:mondscaling} with Eq.~\ref{1djeans} to calculate $\sigma_r(r)$ for a MOND Plummer sphere. The Plummer sphere is generated with $10^5$ particles. We should point out that the Plummer spheres generated by our setup code begin slightly out-of-equilibrium, before quickly relaxing to an equilibrium state, within $3-4$ crossing times in the Newtonian case, and $2-3$ crossing times in the MONDian cases.

\begin{figure}
\centering
\includegraphics[width=8.0cm]{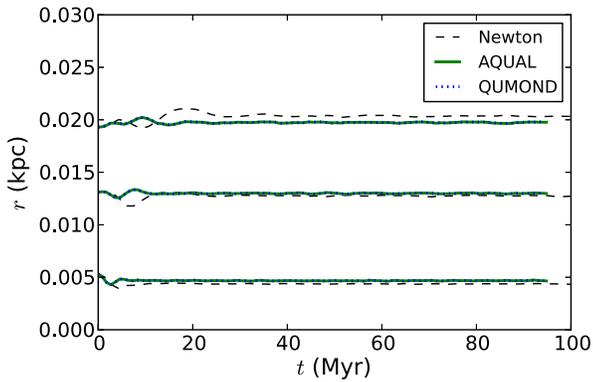}
\caption{The $10\%$, $50\%$ and $70\%$ Lagrange radii of the isolated Plummer spheres, in the Newtonian, QUMOND and AQUAL runs. The two MOND solvers give essentially identical results due to the spherical symmetry of the system.}
\label{isoPlumLR}
\end{figure}

The code is run for a Newtonian system, a QUMOND system and an AQUAL system for approximately $100$~Myr. The Lagrange radii of the Plummer spheres are shown in Fig.~\ref{isoPlumLR}. After initial oscillations all the models quickly settle into equilibrium. Our setup code generates Plummer spheres using a Gaussian distribution of velocities at each radius, while the true velocity distribution in a Plummer sphere is not exactly Gaussian. This approximation leads to initial conditions that are slightly out of equilibrium. For this reason there is some small disagreement between the Lagrange radii for the MONDian and Newtonian systems, however the fact that they are very similar tells us that the density distribution is effectively the same in all three models.

\begin{figure}
\centering
\includegraphics[width=8.0cm]{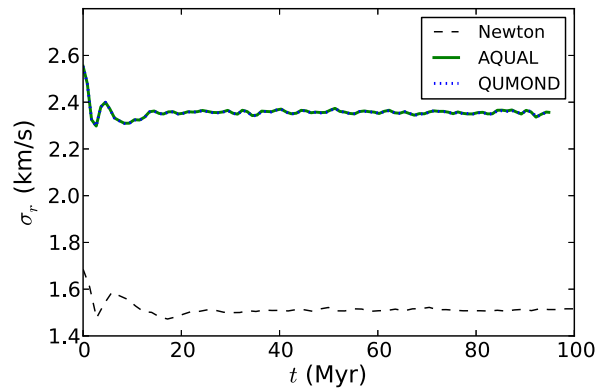}
\caption{Radial velocity dispersion of Newtonian, QUMOND and AQUAL Plummers, at $r=20$~pc. Again the QUMOND and AQUAL lines are identical.}
\label{isoPlumVD}
\end{figure}

The (total) velocity dispersions of the Newtonian and AQUAL systems are shown in Fig.~\ref{isoPlumVD}. The QUMOND result is again essentially identical to that of the AQUAL run. Clearly the MONDian systems have a higher velocity dispersion than the Newtonian system, as expected. This demonstrates that our code is able to produce a stable evolution of a dispersion-supported system in MOND gravity.

\begin{figure}
\centering
\includegraphics[width=8.0cm]{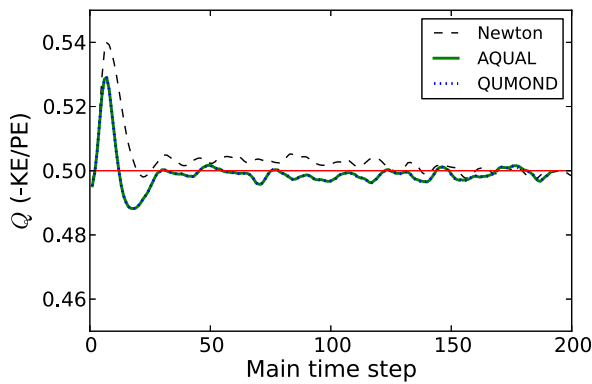}
\caption{Virial ratio for the Newtonian and MONDian Plummers. The Newtonian time step was longer due to slower velocities, so we have continued the Newtonian run to cover the same number of time steps.}
\label{virialratio}
\end{figure}

\begin{figure}
\centering
\includegraphics[width=8.0cm]{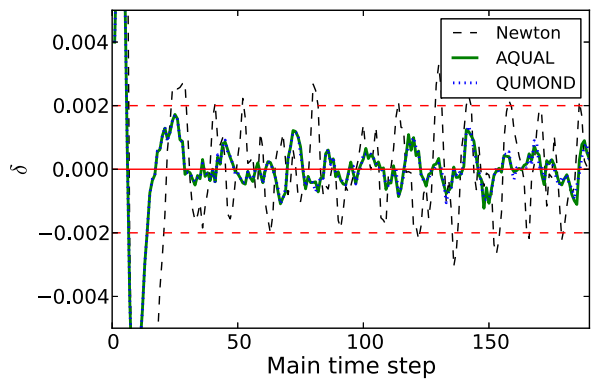}
\caption{Energy conservation parameter for Newtonian, AQUAL and QUMOND Plummers. The energy conservation parameter $\delta$ is defined as $\delta = (E_t-E_{t-1})/E_0$ where $E_t$ is the total energy at time step $t$. We can see that the energy errors (after the initial relaxation of the system) are on the order of $0.2\%$ or less.}
\label{energycons}
\end{figure}

To check energy conservation we follow \cite{PORpaper} and modify the RAMSES code to calculate the potential energy at each time step $t$ as follows:
\begin{equation}
W_t = \sum_{i=1}^{N} \rho_i dx_i^3 (\vec{x}_i \cdot \vec{a}_i)
\end{equation}
where $N$ is the total number of refined cells and, for the $i$-th cell, we have: $dx_i$ is the cell length, $\vec{x}_i$ is the cell position, $\vec{a}_i$ is the acceleration due to gravity (gradient of the potential) in that cell and $\rho_i$ is the density in that cell. The total potential energy is output at each main time step, along with the total kinetic energy $K_t$. The total energy at each main time step is then $E_t = K_t + W_t$. The fractional variation in energy, $\delta$, at each time step is then calculated as
\begin{equation}
\delta = \frac{E_t - E_{t-1}}{E_0}
\end{equation}
where $E_0$ is the total energy at the start of the simulation, $t = 0$. Furthermore, we also calculate the virial ratio at each main time step: $Q_t = -K_t/W_t$. The results for our isolated Plummer sphere simulations are shown in Figs.~\ref{virialratio} and \ref{energycons}. All of the models begin slightly out-of-equilibrium, as already discussed, and quickly settle to equilibrium, $Q = 0.5$. The energy conservation appears satisfactory in all runs, with energy errors on the order of $0.2\%$ in the Newtonian case and slightly \emph{less} in the MONDian runs. The reason for this behaviour is because of the larger total energies involved in the MOND simulations. The standard deviation of the total energy fluctuations are comparable in both the Newtonian and MONDian runs, but when we normalise with the total energy we find a slightly smaller fractional error for the MOND runs.

It is worth noting that the fractional energy conservation errors for the QUMOND POR code described in \cite{PORpaper} are larger in the MOND case than in the Newtonian case, while our code shows the opposite behaviour: the MOND fractional errors are \emph{smaller} than the Newtonian fractional errors. This is likely due to the different numerical implementation (the stencil used in the POR code differs from our stencil, for example) although the comparison is not exact as the test run used in \cite{PORpaper} is at considerably higher resolution than the test case we have shown here.

\subsubsection{Orbiting Plummer spheres}
\label{subsec:tests_orbPlummer}

\begin{figure*}
\begin{tabular}{cc}
  \includegraphics[width=7.5cm]{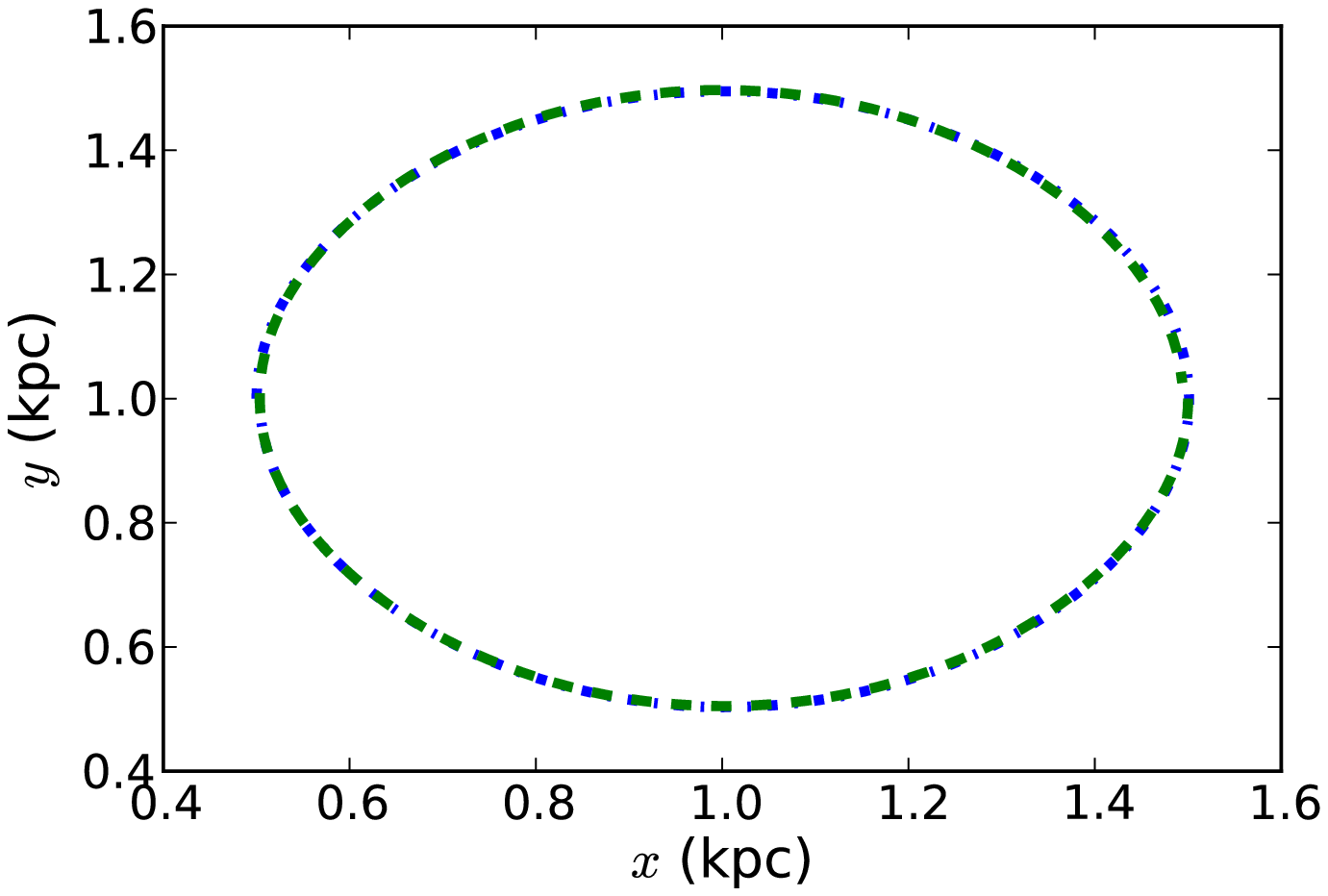} &   \includegraphics[width=7.5cm]{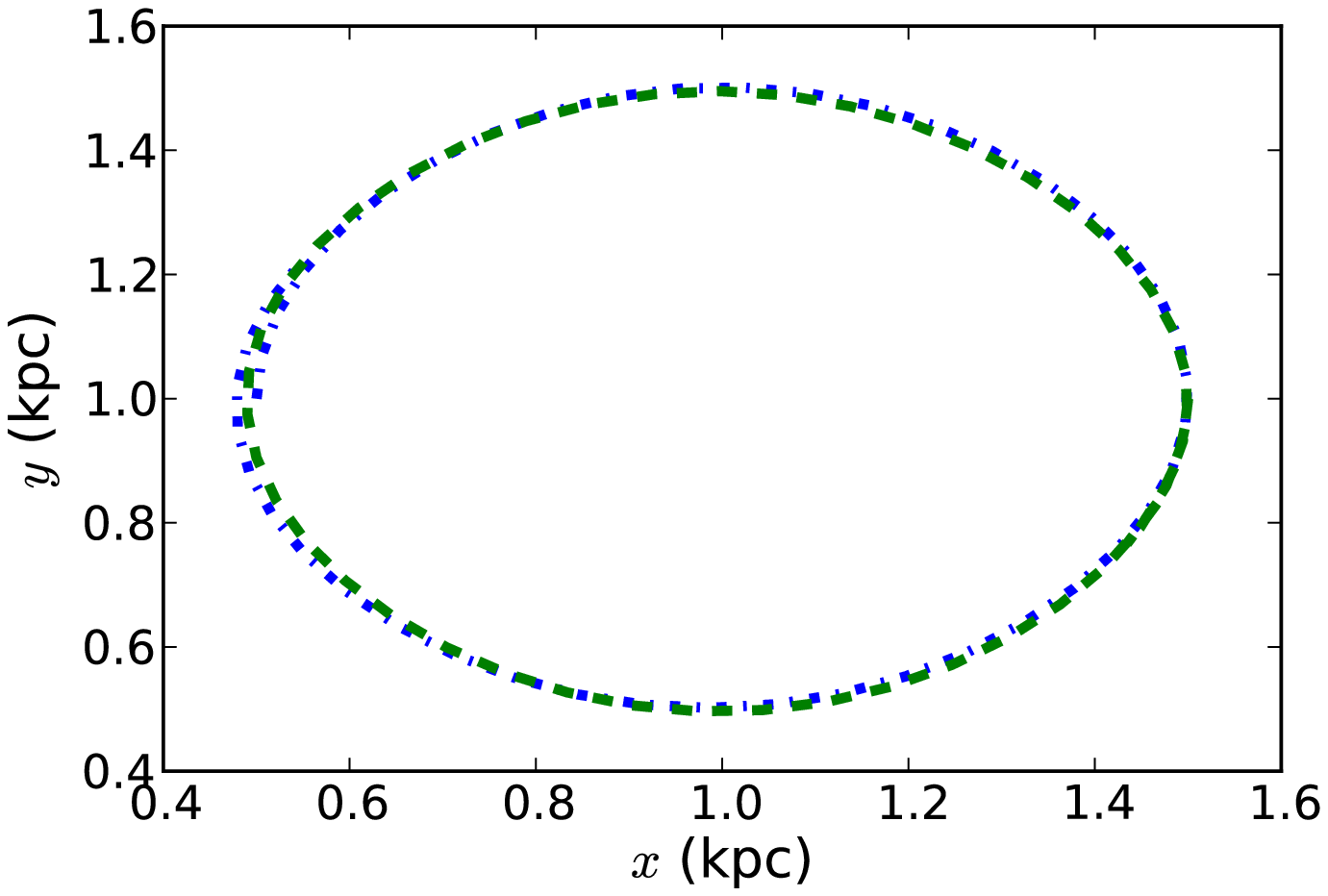} 
\end{tabular}
\caption{Centre-of-mass trajectories for 2 orbiting Plummer spheres, Newtonian case on the left, AQUAL on the right. The trajectory of one Plummer sphere is marked with a green dashed line, and the other with a blue dot-dashed line.}
\label{orbitsCoM}
\end{figure*}

\begin{figure*}
\begin{tabular}{cc}
  \includegraphics[width=7.5cm]{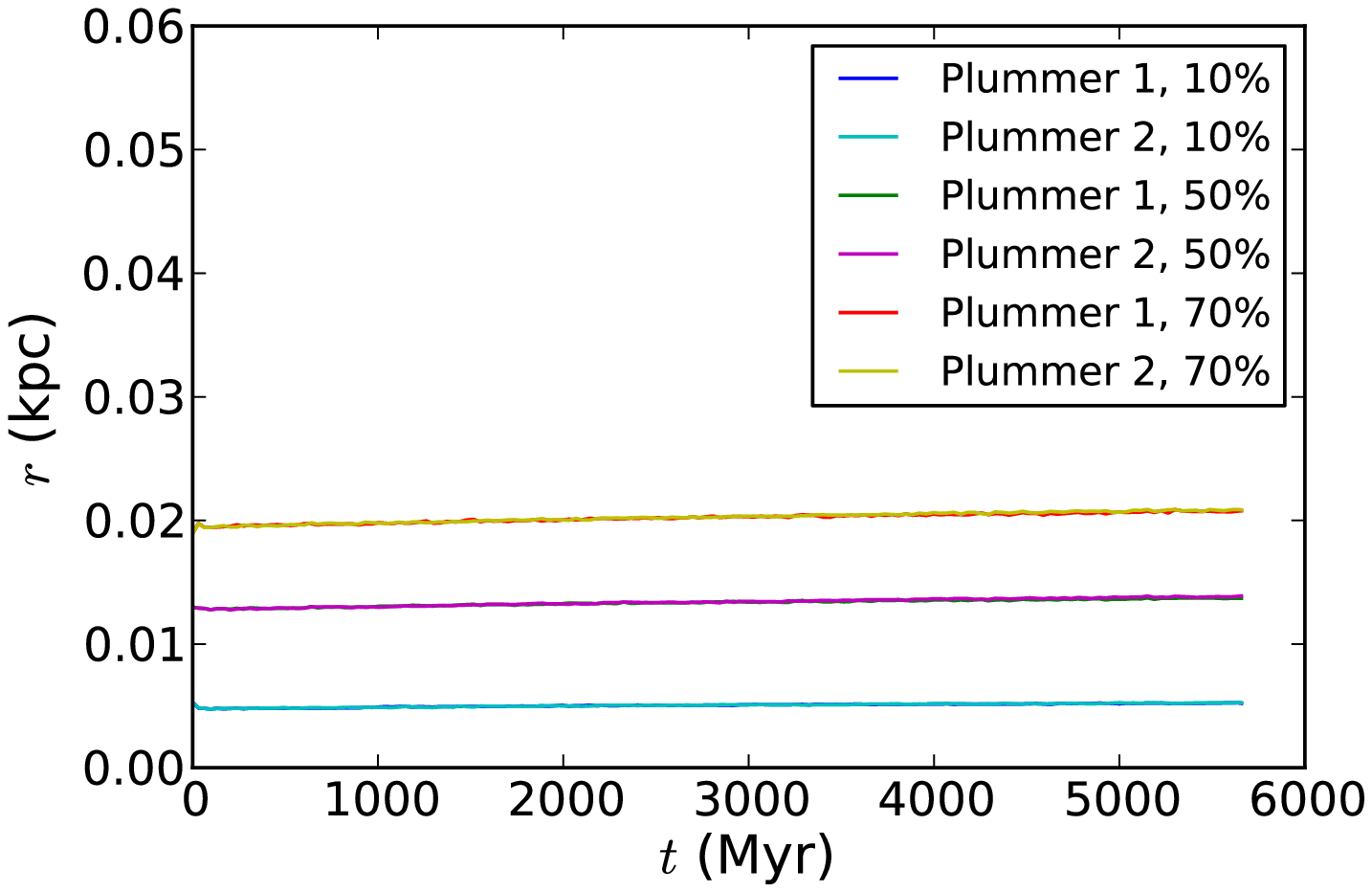} &   \includegraphics[width=7.5cm]{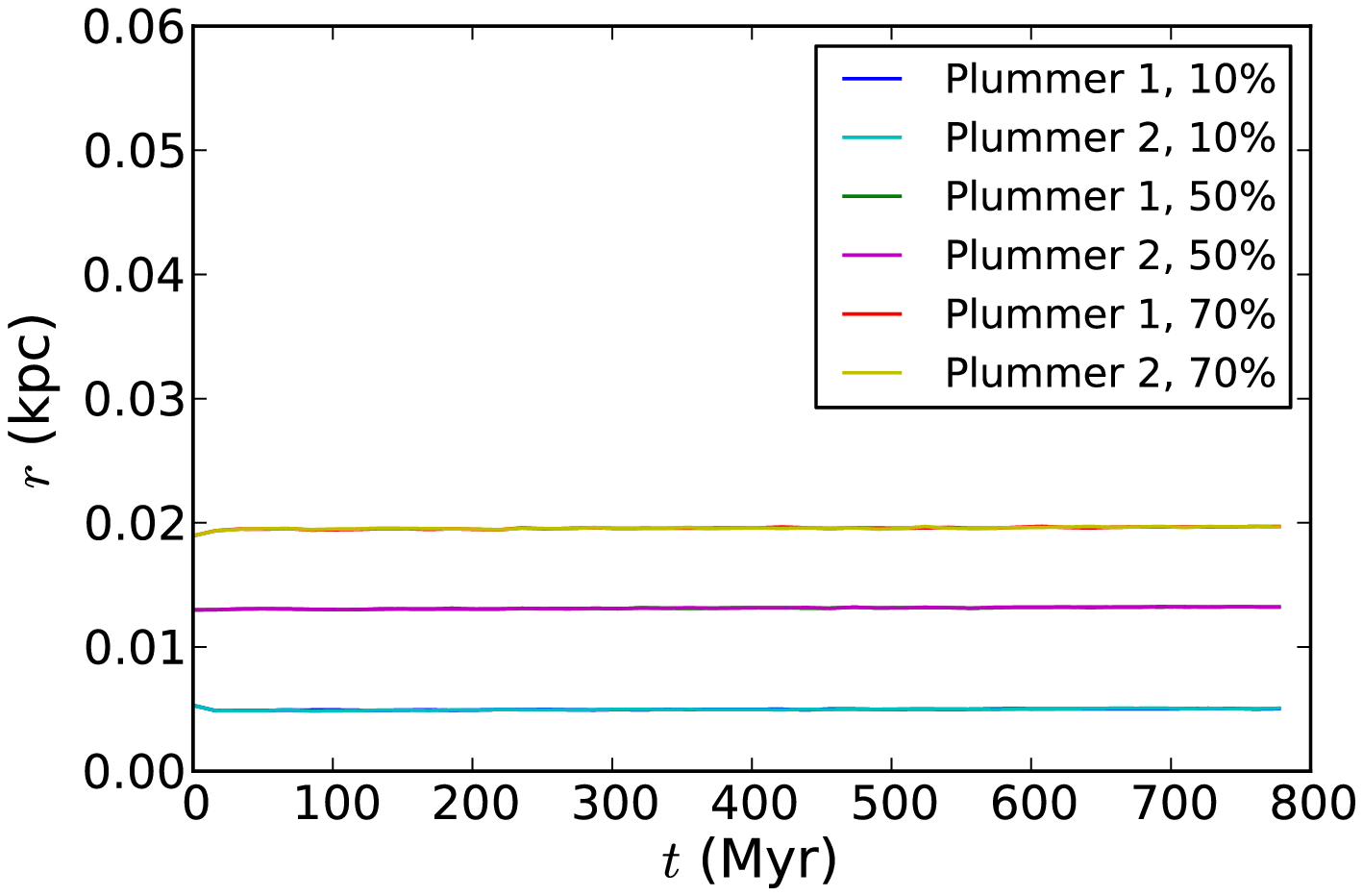}
\end{tabular}
\caption{Lagrange radii for 2 orbiting Plummer spheres, Newtonian case on the left, AQUAL on the right.}
\label{LRorbitPlum}
\end{figure*}

After discussing isolated Plummer spheres in the previous section, we now wish to demonstrate the use of the code to evolve a system of two orbiting Plummer spheres. The two-body problem is, of course, analytically solvable in Newtonian gravity, but this is not the case in a MONDian system, except in the special case of circular orbits in the deep MOND regime (\citealp{zhaotwobody}). We set up two equal mass Plummer spheres, with the parameters given in the previous Section, and, by trial and error, give them tangential velocities to put them in approximately circular orbits around each other. We placed the Plummer spheres at $x= \mp 0.5$~kpc in the simulation box (they are centred in the $y$ and $z$ directions) with velocities in the $y$ direction of $\pm 0.52$~km/s for a comparison Newtonian case, and $\pm 4.15$~km/s in the MONDian cases. To speed up the orbital integration we chose to run these simulations on lower resolution meshes, compared with those used in the isolated Plummer simulations. The highest level of refinement in these runs was level 9 (a maximum resolution of $1.95$~pc).

The orbits of the centre-of-masses of the Plummer spheres are shown in Fig.~\ref{orbitsCoM} for the Newtonian run and the AQUAL run. The Newtonian simulation was run for approximately $5.7$~Gyr, while the AQUAL simulation was run for a shorter time of approximately $770$~Myr. The orbits are very close to circular in both cases, with a stronger deviation from circularity in the AQUAL case, due to using a slightly incorrect initial velocity. In both cases the two Plummer spheres have completed slightly more than a whole orbit, returning to close to their starting positions. We can see, therefore, that the orbital velocity in the MOND case is much larger than the Newtonian case, as the Newtonian Plummer spheres take over $5$~Gyr to complete an orbit, while the AQUAL Plummer spheres have done this in only around $700$~Myr.

The Lagrange radii of the two orbiting Plummer spheres are shown in Fig.~\ref{LRorbitPlum}. As these simulations were run with slightly lower resolution in order to speed up the orbital integration, the Plummer spheres are not quite as stable as in the isolated case: there is a very gradual expansion of the Plummer spheres over time, which is clearer in the Newtonian case because of the much longer timescale of the simulation. These test simulations demonstrate that our code is able to evolve a stellar system in an orbit around another system, either in the Newtonian or MONDian regimes. This suggests that our code is conserving momentum, and the fact that the Plummer spheres are essentially stable implies that we do not have any spurious numerical issues affecting the internal dynamics. Furthermore, this provides an example of the usefulness of adaptive mesh refinement: we can model stable, compact stellar systems orbiting in a background potential, where we only need to apply the high resolution mesh to the orbiting systems. Such configurations correspond to simulations of satellite galaxies orbiting their host, or large spirals in clusters.

\subsubsection{Full galaxy model}
\label{subsec:tests_expdisks}
We now move on to N-body tests of a galaxy model using both a disk and a bulge component. We use the initial conditions code \emph{GalactICS} (\citealp{galactics}) to generate a Newtonian disk and bulge inside a dark matter halo. We run a simulation of this model as a control run using the normal Newtonian RAMSES code. For the MOND runs, we simply remove the dark matter halo particles from the initial conditions and run the disk and bulge particles only using RAyMOND. Fortuitously, this seems to result in a disk with a very similar rotation curve. Thinking of this in terms of a ``phantom dark matter'' (PDM) halo, it appears that in this case the Newtonian halo corresponds closely with the PDM halo, but this may not generally be true.

The parameters in the disk, bulge and halo are those of the MW-A model in \cite{galactics}. The disk is modeled as an exponential disk, with mass $4.2 \times 10^{10} M_{\odot}$, scale radius $4.5$~kpc, and scale height $0.43$~kpc. The bulge is a King model, with mass $2.1 \times 10^{10} M_{\odot}$ and the halo uses a lowered Evans model with mass $2.7 \times 10^{11} M_{\odot}$ (see \cite{galactics} for the full set of parameters used to initialise the model). We use $60000$ particles in the disk, $30000$ in the halo and $10000$ in the bulge.

\begin{figure*}
\begin{tabular}{cc}
\includegraphics[width=7.5cm]{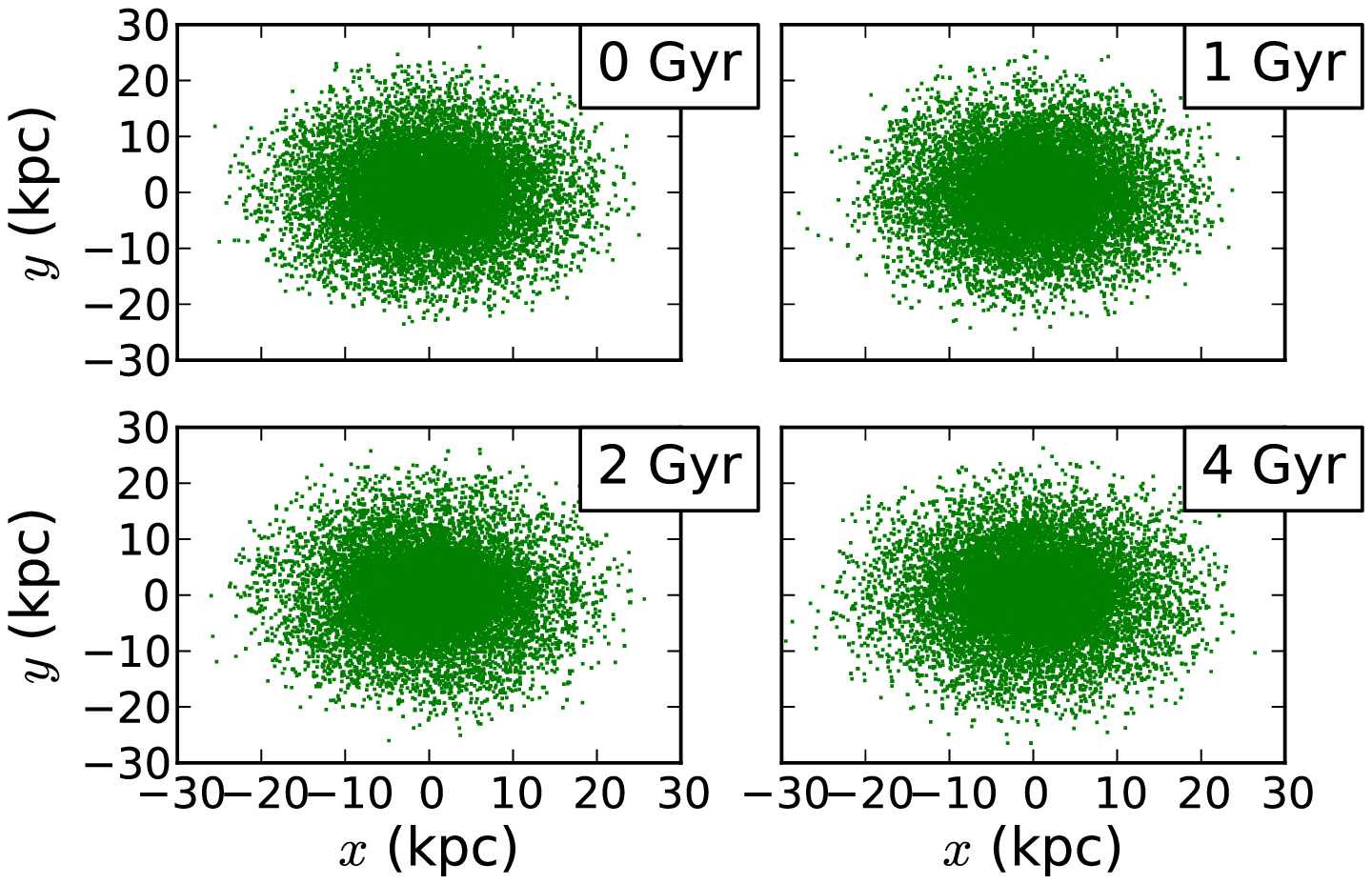} & \includegraphics[width=7.5cm]{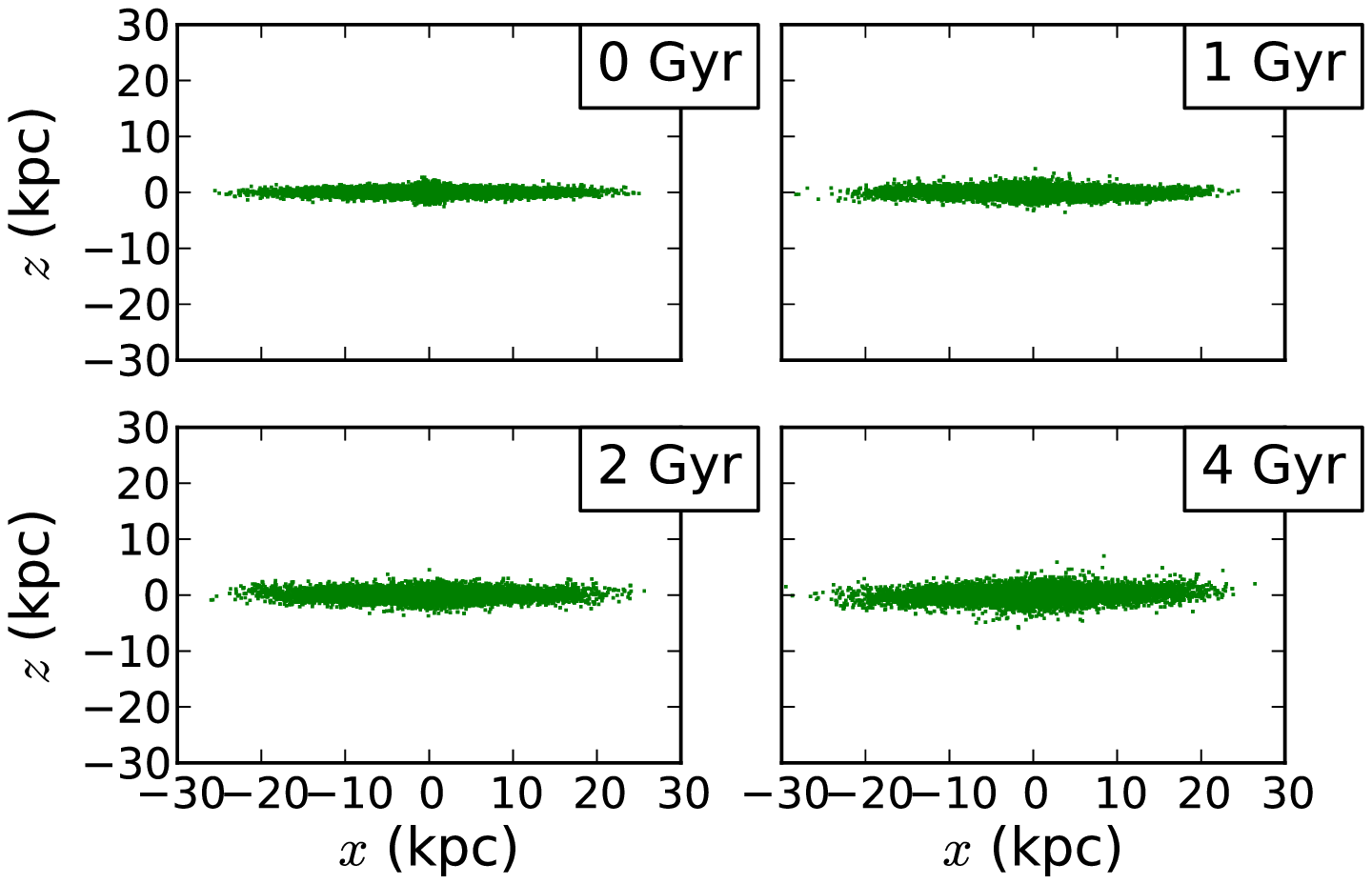}
\end{tabular}
\caption{Newtonian disk and bulge (dark matter halo is not plotted). These are subsets of $20000$ particles.}
\label{newton_disk}
\end{figure*}

\begin{figure*}
\begin{tabular}{cc}
\includegraphics[width=7.5cm]{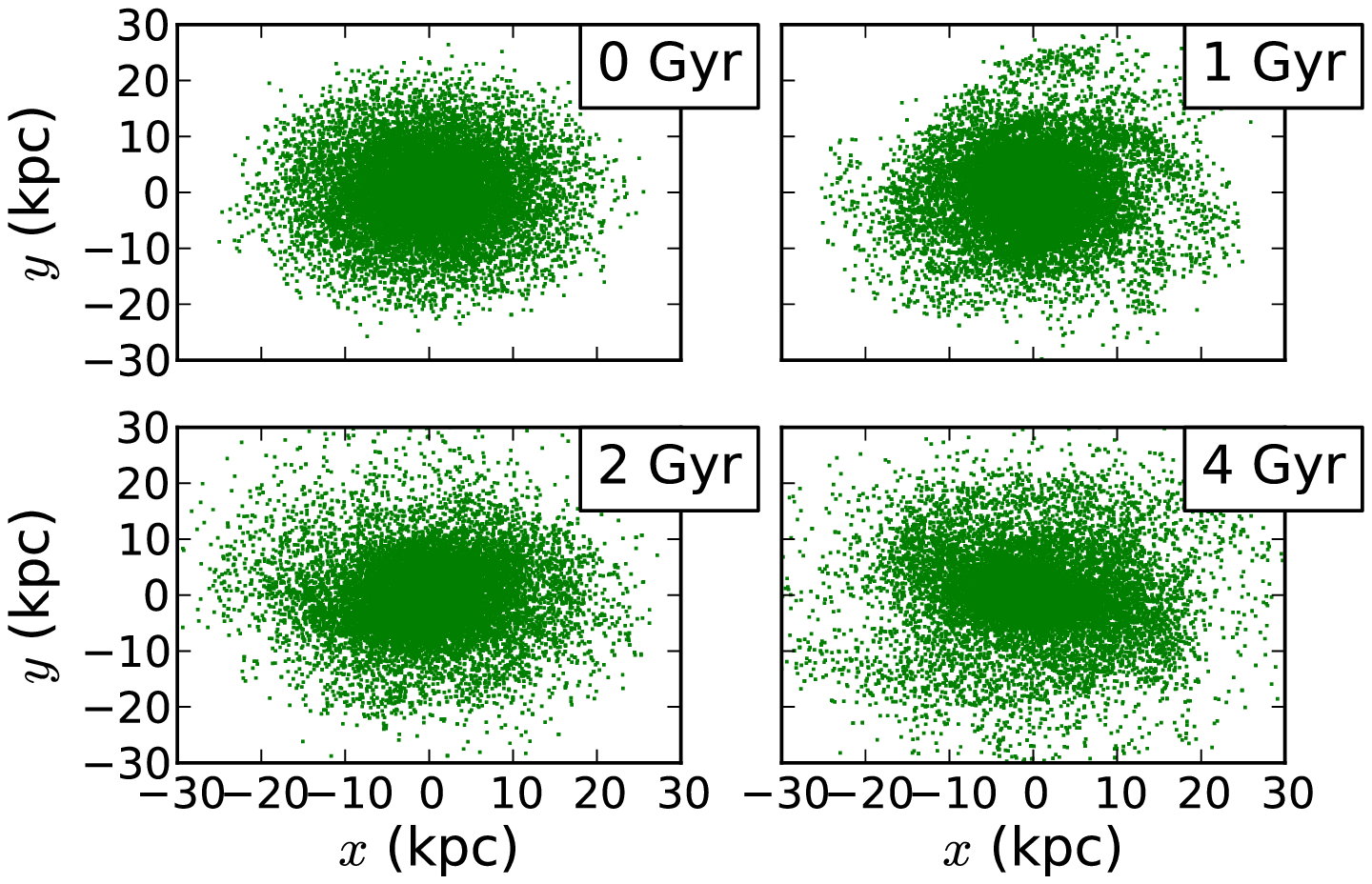} & \includegraphics[width=7.5cm]{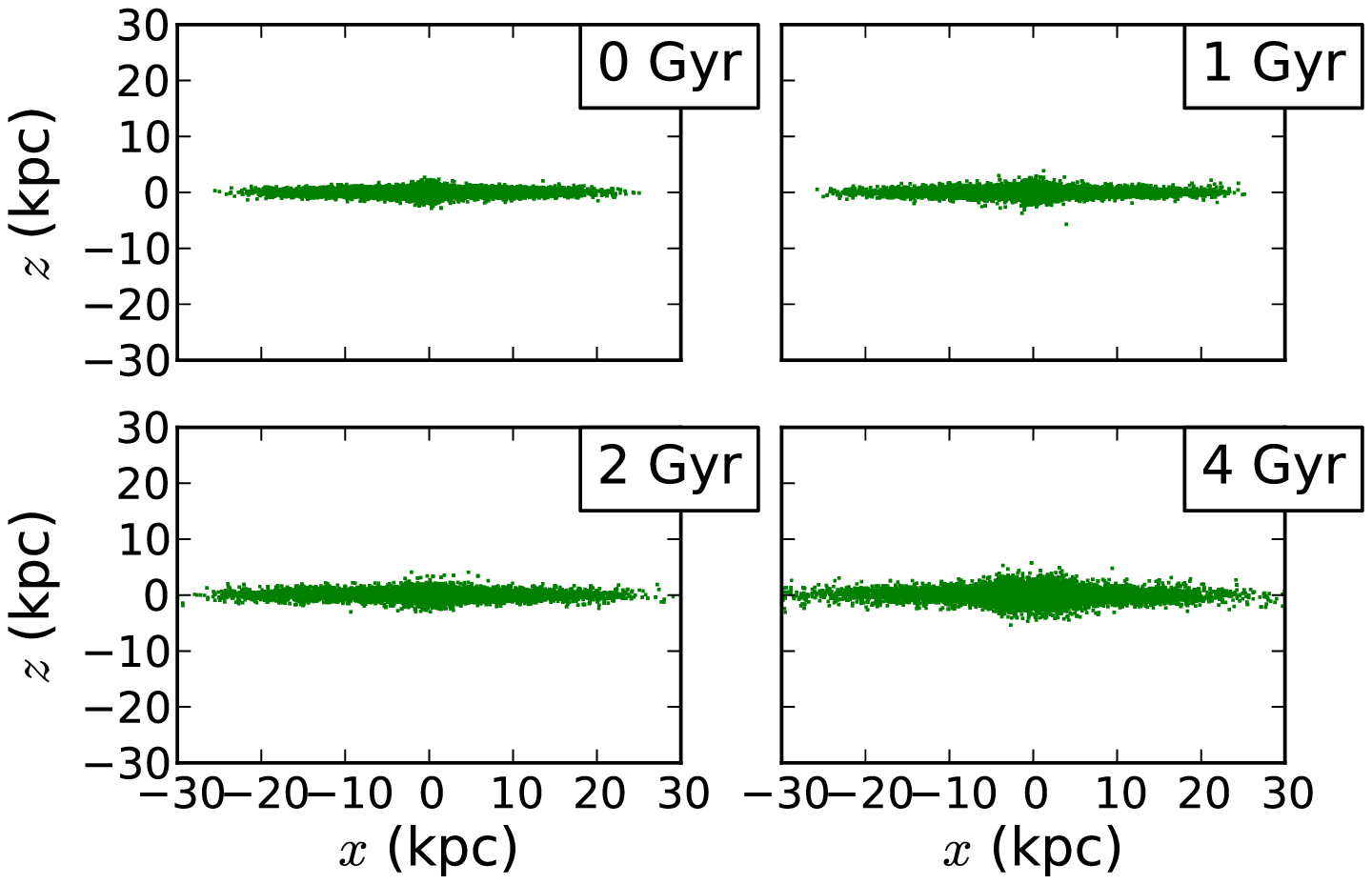}
\end{tabular}
\caption{AQUAL disk and bulge (no dark matter halo). These are subsets of $20000$ particles.}
\label{aqual_disk}
\end{figure*}

Snapshots of the Newtonian run viewed face-on and edge on are shown in Fig.~\ref{newton_disk}. Although not clear in the plots, the disk forms weak spiral structure soon after the simulation starts, but is clearly globally stable, as expected due to the presence of the dark matter halo. There is some disk thickening over time, as is standard in numerical disk simulations. The AQUAL run snapshots are shown in Fig.~\ref{aqual_disk}. The disk forms more pronounced spiral structure than in the Newtonian case, but again exhibits no global instability. A bar instability does develop after around $3$~Gyr of evolution. The disk thickening is considerably less pronounced in this case, although the bar becomes clearly visible by $4$~Gyr.



\begin{figure}
\centering
\includegraphics[width=8.0cm]{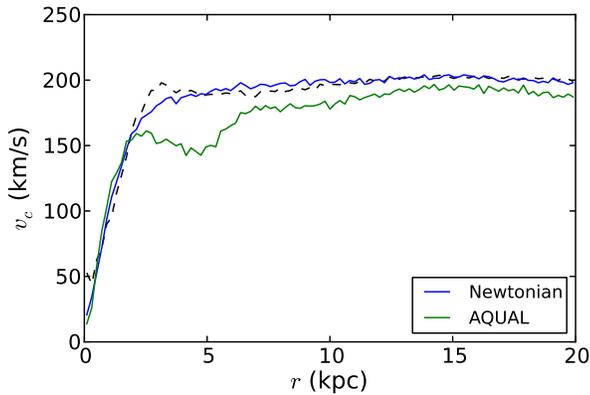}
\caption{Mean rotational (azimuthal) velocity of the Newtonian disk and bulge and the AQUAL disk and bulge at the beginning and end of their respective runs, in $r$-bins of width $200$~pc. The black line denotes the initial conditions of the simulations, the blue dashed line shows the Newtonian rotation curve after approximately $4$~Gyr, while the green solid line shows the AQUAL rotation curve after the same time.}
\label{meanv_comparison}
\end{figure}

The average rotational velocities as a function of radius in the disk are shown in Fig.~\ref{meanv_comparison}. The rotation curve at the beginning of the simulation (identical in both the Newtonian and AQUAL runs) is shown as a black line, the Newtonian rotation curve after $4$~Gyr is shown as a dashed blue line, while the AQUAL rotation curve is shown as a solid green line. In the Newtonian case, the flat rotation curve is evident due to the presence of the dark matter halo, and it remains stable throughout the evolution. The AQUAL rotation curve remains effectively flat at large radii, entirely due to the MOND enhancement of the gravitational accelerations in the disk, as there is no dark matter halo in this simulation. The MOND curve is slightly lower than that of the Newtonian system after $4$~Gyr, and the existence of the bar is evident in the dip in the rotation curve at small radii.

This test demonstrates that our code can evolve a rotationally-supported system over a significant timescale, and that our code may be used in the study of secular processes in MOND disks.

\subsubsection{Comparing AQUAL and QUMOND}
\label{compaqualqumond}
Although the differences between the two formulations of MOND utilised in our code are expected to be small, the differences that would build up over time between a QUMOND and an AQUAL system may well be detectable.

As a basic illustration of the fact that the two formulations differ, we place an analytic Plummer profile (again with $r_{pl} = 10$~pc and $M_{pl} = 10^5 M_{\odot}$) in the background field of an analytic exponential disk model with a bulge. This disk has $M_d = 5.4 \times 10^9 M_{\odot}$, a scale radius of $2.3$~kpc, and a scale height of $0.23$~kpc. The bulge is modelled by a Hernquist sphere with $M_b = 1.68 \times 10^9 M_{\odot}$ and scale radius $0.5$~kpc. We place the Plummer sphere at the origin of our computational domain, while the centre of the bulge of the disk galaxy is located at $\vec{r} = (10/\sqrt{3},10/\sqrt{3},10/\sqrt{3})$, i.e. at $10$~kpc ``diagonally'' away from the Plummer sphere. In this way we can examine the behaviour of the external field effect on the internal accelerations of the Plummer sphere. As stated in the Introduction, the external field effect is known to differ in these two formulations of MOND.

\begin{figure}
\centering
\includegraphics[width=8.0cm]{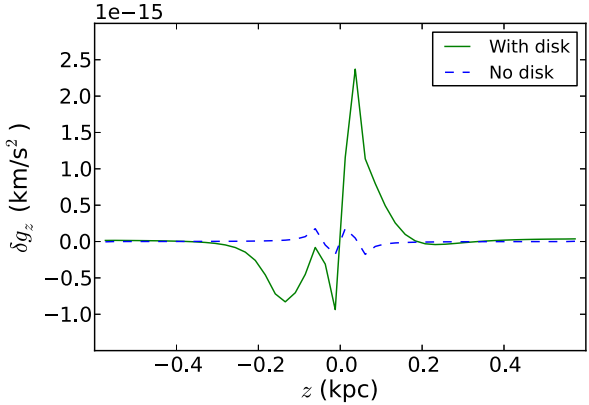}
\caption{The difference between the vertical accelerations near the centre of the Plummer sphere in the two MOND formulations: $g_z^{\text{AQUAL}} - g_z^{\text{QUMOND}}$. The blue dashed line shows the differences when the background galaxy model is not present, while the solid green line shows the differences in the presence of the background galaxy model.}
\label{compare_aqual_qumond}
\end{figure}

The accelerations throughout the system, at varying levels of refinement, are determined using a single time-step calculation of RAyMOND with a box length of $50$~kpc. We consider the difference in the AQUAL and QUMOND accelerations along the $z$ direction:
\begin{equation}
\delta g_z = g_z^{\text{AQUAL}} - g_z^{\text{QUMOND}},
\end{equation}
with $x, y \approx 1.2$~pc away from the origin (this is because the computational mesh only samples certain points in the domain). The differences at refinement level 11, corresponding to a cell length of $2.4$~pc, are shown in Fig.~\ref{compare_aqual_qumond}. The blue dashed line shows the differences when the background disk model is not present, and so there is no external field effect. There is a small difference in the accelerations at the centre of the Plummer sphere, due to errors arising from the finite resolution of the grid. When the background galaxy model is present, however, the differences become somewhat larger. Although a difference in the accelerations of $\sim 10^{-15}$ km/s$^2$ is rather small, this is on the order of $1\%$ of the MOND acceleration scale $a_0$. Furthermore, the cumulative effect of such differences over significant timescales (such as during a satellite orbit over several Gyr) may well become substantial. Finally, it is possible that in other systems the differences between the two formulations is larger than our purely illustrative example shown here. We leave the investigation of such differences for future work.

\subsection{Execution time analysis}
\label{timing}
Due to the extended stencil used in the iterative solver in the AQUAL code, there are $18$ neighbouring grid points as opposed to $6$ in standard RAMSES. Furthermore, the solver must evaluate the $6$ neighbouring $\mu$ values. In total, the AQUAL code performs calculations at $24$ points for each sweep through the iterative solver, and thus we would expect that this code is at least four times slower than the Newtonian code. The QUMOND code also uses an extended stencil, but this is only required for one pass through the mesh, when the additional density contribution is calculated. The main reason for slower execution of the QUMOND code is because we run the solver twice each time step. Therefore, we expect that the QUMOND code runs at least two times slower than the Newtonian code.

\begin{figure}
\centering
\includegraphics[width=8.0cm]{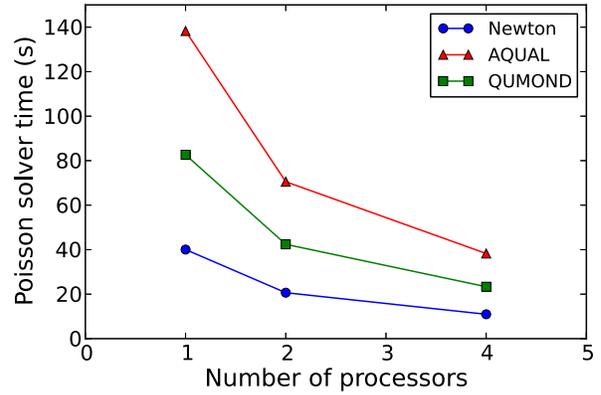}
\caption{Code timings for the isolated Plummer N-body simulations. These are the average times between two main time steps for varying numbers of processors.}
\label{timings}
\end{figure}

\begin{figure}
\centering
\includegraphics[width=8.0cm]{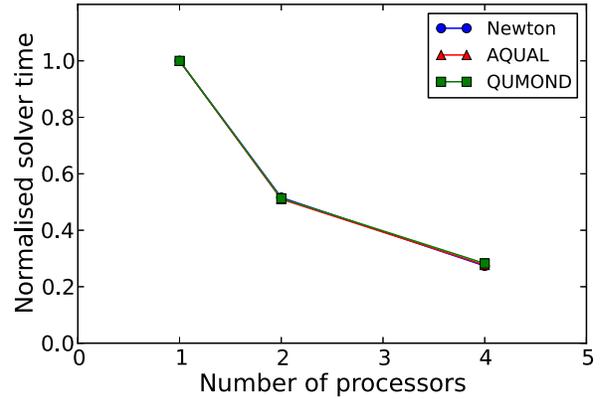}
\caption{The code timings of Fig.~\ref{timings} normalised by the time taken for one processor in each case. The parallelisation speed-up is essentially identical for all the versions of the code.}
\label{timings_norm}
\end{figure}

We perform a simple timing analysis by running the isolated Plummer model described in Section~\ref{subsec:tests_isoPlummer} for the Newtonian, AQUAL and QUMOND codes, as serial runs, and as parallel runs on 2 processors and on 4 processors. RAMSES calculates the time elapsed between each main time step, i.e. whenever the code returns to the base level mesh that covers the entire computational domain. We use this time to compare the codes. The absolute timing results are given in Fig.~\ref{timings}, while in Fig.~\ref{timings_norm} we have normalised the times by the time taken for a single processor. This allows us to check that the speed-up from parallelisation is not adversely affected by our modifications.

As expected, the AQUAL code is, on average, about four times slower than the Newtonian code, while the QUMOND code is about two times slower. The parallelisation speed-up is identical in all three cases, showing that our modifications for the MOND calculation have had no adverse effect.

\section{Summary and conclusions}
\label{concl}
We have developed a modification of the N-body/hydrodynamics code RAMSES that enables high resolution simulations using MOND gravity. The code utilises two formulations of MOND, known as AQUAL and QUMOND, allowing a direct comparison between them and the investigation of possible observational consequences from the different manifestations of MONDian behaviour. The execution time of the code has also been demonstrated to be as fast as may be expected, considering the complexity of the MOND calculation. Furthermore, using idealised tests we have confirmed that the RAyMOND code is able to evolve dispersion-supported and rotation-supported stellar systems in a stable manner in a MOND gravitational potential. The code will soon be available on request and a freely available public release is planned for the future.

One of the principal advantages of using RAMSES for this work is that the modular design of the code allowed us to modify only the gravitational solver: we did not have to make any adjustments to the hydrodynamics solver. In addition, the hydrodynamics solver simply uses whatever gravitational potential is produced by that part of the code, meaning that our code is capable of hydrodynamics simulations using MOND gravity.

As stated in the Introduction, the primary motivation for the development of this code was the study of how MOND may affect the dynamics of galaxies in different environments, and how the physical processes affecting galaxies in those environments may be modified. A specific example of this is the study of the consequences of the external field effect, and how a difference between the EFE in the two formulations of MOND may lead to observational differences between MOND theories, as well as standard $\Lambda$CDM predictions. Beyond the applications to galaxy dynamics, the code may also be used to further investigate the consequences of MOND for cosmology. While still not a fully consistent treatment for cosmology (there is no relativistic formulation of MOND implemented for the behaviour of the Hubble flow) it may nonetheless prove useful to compare and contrast the development of structure formation in the two MOND formulations.

There are, of course, many possible applications of our code at galactic scales, including galaxy interactions, the behaviour of dwarf satellites, star cluster dynamics in the Milky-Way and so on. Such studies of the effects of MOND gravity on galaxy dynamics in various environments will be the subject of a series of future publications.

Through applications of MOND gravity to more complex circumstances of galaxy interactions and environmental effects, we hope to more fully explore the consequences of this persistent idea. It may well be that within the domain of applicability where MOND has historically performed very well, there are possibilities of finding conclusive observational signatures of MOND that contrast markedly to those of $\Lambda$CDM, and ultimately a route to improve our understanding of dark matter, modified gravity, or both.

\section*{Acknowledgments}
The authors wish to thank Romain Teyssier for providing very helpful information about RAMSES via the RAMSES users mailing list. They would also like to thank Francoise Combes for suggesting the use of RAMSES for this project, and Marcel Pawlowski for additional comments. Finally we wish to thank the anonymous reviewer for helpful comments and suggestions. GC, RS and MF acknowledge the support of FONDECYT grants 3130480, 3120135 and 1130521 respectively.


\end{document}